# Species Population Dynamics with Competition and Random Events

M. M. Tehrani and S. Soltanieh

Innova Photonics, Westlake Village, CA. USA

## Abstract

Population dynamics of a competitive two-species system under the influence of random events are analyzed and expressions for the steady-state population mean, fluctuations, and cross-correlation of the two species are presented. It is shown that random events cause the population mean of the both species to make smooth transition from far above to far below of its growth rate threshold. At the same time, the population mean of the weaker specie never reaches the extinction point. It is also shown that, as a result of competition, the relative population fluctuations do not die out as the growth rates of both species are raised far above their respective thresholds. This behavior is most remarkable at the maximum competition where the weaker specie's population statistics becomes completely chaotic regardless of how far its growth rate is raised.

## I. Introduction

Mathematical models of population ecology have evolved from the simple Exponential Law of Thomas Malthus [1] to the Logistic Equation of Quetelet and Verhulst [2] to the Lotka-Volterra model that includes competition [3]-[4] and, finally, to models that consider the effects of random events [5]-[8]. Various mathematical methods have been employed in models that include random processes. Amongst them is the Fokker-Planck approach that provides for the time evolution of probability density as a function of system dynamics and diffusion forces [9]-[11]. In general, it is assumed that after a sufficiently long time the system reaches a stationary state and the probability density becomes time independent. Thus, the steady-state solution of the Fokker-Planck equation is obtained and is used to derive expressions for the population mean and other statistical quantities using various approximations.

In studying the behavior of two-mode ring lasers, one of us (MMT) along with L. Mandel developed the Coherence Theory of the Ring Laser and published the results in a number of papers [12]. In such a system, the two counter-propagating waves (modes) of the laser gain their optical power from the photons emitted from the excited states of the atoms in the gain medium. As such, they also compete with each other for the emitted photons and the competition becomes stronger as the operating frequency of the laser gets closer to the center of the gain medium's emission line. At that point, both modes draw their photons from the same set of atoms. In addition to the "coherent" photons that the gain medium supplies to each mode, it also injects spontaneously emitted photons that act as noise in each mode. In [12], we laid out the mathematical foundation for this process. Using the Fokker-Planck technique, we obtained the probability density function of the two modes as a function of each mode's amplitude and their degree of competition. We used the steady-state probability density function to derive equations for the intensity mean and fluctuations of each mode. Predictions of this theory were verified in a series of experiments [13].

There is a striking similarity between the behavior of a two-mode laser and that of two entities competing for the same resources. If one substitutes the resource for which the two entities compete with photons, one arrives at the same scenario as the two-mode laser. In fact, the mathematical formulation presented in [12] is identical to those discussed in population ecology that include random events [9]-[11]. Of course, there are vast differences between the complexity, variety, and time scales of the two systems. For example, the device parameters in the two-mode laser can be set and controlled with a high degree of accuracy and their exact values and effects on device performance are readily predictable. However, this is not the case in the parameters that determine the two-entity dynamics and one has to either guess or rely on a large set of past data that may have been obtained in different environments. Also, while the statistical properties of the spontaneous emission noise that affect the two-mode laser is well known, those affecting ecological systems have multitude of origins each with its own (mostly unknown) statistics. Nevertheless, the model presented here can at least provide a qualitative prescription for effects of system parameters on its dynamics.

In this paper, we present a formulation of the dynamics of two species competing for the same resources. Competition can be Interaspecific, Interspecific, Interfering, or Exploitative. We allow each specie to have its own growth rate but with a symmetric competition coefficient. Our formulation, adapted from the Coherence Theory of the Ring Laser [12], includes random events that affect the population dynamics of each specie. Using the Central Limit Theorem, we assume the totality of stochastic processes, environmental and demographic add up to a Gaussian random process. We use the machinery of Fokker-Planck equation and derive exact analytical solution for the steady-state probability density function that we use to obtain various moments of the distribution. In particular, we calculate the first and second order moments that represent the mean and the fluctuations of each specie's population, respectively. As it has been pointed out [10, Chapter 5], the steady-state probability distribution is to the stochastic environment as the stable equilibrium point is to the deterministic case of Lotka-Volterra. In our analysis, the first and second order moments represent the equilibrium point and its fluctuations, respectively. A significant result of our model is that, in the absence of a dominating and high impact random event, stochastic processes allow for non-zero population mean of specie even if its growth rate is at or below zero. This result is accompanied by an increase in population fluctuations.

We set up the mathematical framework in Section II by starting with Lotka-Volterra equations and, in Section III we add a random function to each equation. We solve the resulting Fokker-Planck equation and obtain its steady-state solution in Section IV and use it in Section V to derive various moments of the distribution. In particular, we focus on the first two moments that yield each specie's population mean, its fluctuations and their cross correlations. In Section VI, we discuss special cases of 1): two species at their growth rate threshold, 2): two species at their maximum competition, and 3): two species at zero competition. The last case signifies the population dynamics of single specie under the influence of random events. Our conclusions and some suggestions for future work are presented in Section VII. In this paper, we will not duplicate the detailed derivations of the equations that are presented in Ref. 12 and to which we refer the interested readers.

## II. Theoretical Modeling

We start with the equations for the rate of change of populations of a competitive two-species system. They can be written as:

$$\frac{d}{dt}N_1 = a_1 N_1 - N_1^2 - \xi N_1 N_2 \tag{1}$$

$$\frac{d}{dt}N_2 = a_2 N_2 - N_2^2 - \xi N_2 N_1 \tag{2}$$

Although borrowed directly from the dynamics of an interacting two-mode laser [12], Eqs. (1) and (2) are known as Lotka-Volterra equations in population ecology and describe the population dynamics of two competing species [3]-[4].

In the above equations, $N_1$ and $N_2$ represent populations or population densities of the two species in a given period. Other parameters are defined below (with indices 1 and 2 referring to the two species):

$a$ is the normalized growth rate of each specie and is the difference between the gain and loss of the population divided by the saturation coefficient. Birth and joining of two identical species into one territory are amongst the sources of growth rate. Loss is the accumulation of all *known* mechanisms that decrease the population in the same period. These include natural death, predation, internal fighting, etc. Each of the components of gain and loss may have different frequencies and impacts on the specie's population. However, we lump all of them into one parameter. This is a reasonable approach if the observation time for population change is long compared to the periods of all mechanisms involved. In general, the growth rate is a combination of a deterministic or predictable term and a random process (demographic stochasticity). Here we take $a$ to be the deterministic portion of the growth rate and will address its random component when we add random functions to (1) and (2). The case of $a=0$ is called the growth rate threshold.

$\xi$ is the competition coefficient and is numerically positive. We assume that the competition coefficient is symmetric between the two species. The competition effectiveness on the population of the two species depends on the population of each and the value of the competition coefficient. We take the coefficient to be between 0 (no competition) and 1 (maximum competition). The condition $0 \leq \xi \leq 1$ is consistent with the Gause-Lotka-Volterra criterion for a stable two-species competing system [5].

In Eqs. (1) and (2), the second term on the right hand side ($N_1^2$ and $N_2^2$) represents the population saturation or carrying capacity due to resource limitations. Without it, the populations grow exponentially and follow the logistic growth process [2]. It might be considered a loss mechanism whose magnitude depends on the size of the population. Typically, this term has its own coefficient. However, for convenience, we scale $a$ and $\xi$ to the saturation coefficient and show the term as $-N_{1,2}^2$. The third term represents the competition from the other specie whose magnitude depends on the size of the competing population and the competition coefficient $\xi$.

The steady-state solutions of Eqs. (1) and (2) are found to be:

$$N_1(s) = \frac{a_1 - \xi a_2}{1 - \xi^2} \quad \text{and} \quad N_2(s) = \frac{a_2 - \xi a_1}{1 - \xi^2} \tag{3}$$

It is seen from Eqs. (3) that $N_1(s) > N_2(s)$ if $a_1 > a_2$ (and vice versa) for all values of the competition coefficient $\xi$. Also, Eqs. (3) become indeterminate for $\xi = 1$ if the growth rate of the two species are different. For equal rates ($a_1 = a_2 = a$) the two populations become equal and given by

$$N_1(s) = N_2(s) = \frac{a}{1 + \xi} \tag{4}$$

In the absence of any competition ($\xi = 0$), each specie's population changes linearly with its growth rate and the specie becomes extinct at or below threshold. We will show that these results change drastically when we include random events.

## III. Langevin and Fokker-Planck Equations

To include random processes that affect the population dynamics of the two species system, the deterministic Lotka-Volterra equations (1) and (2) need to be augmented with random functions that mimic the underlying processes. This can be accomplished by adding two independent random functions $q_1(t)$ and $q_2(t)$ to Eqs. (1) and (2) to obtain:

$$\frac{d}{dt} N_1 = a_1 N_1 - N_1^2 - \xi N_1 N_2 + q_1(t) \tag{5}$$

and

$$\frac{d}{dt} N_2 = a_2 N_2 - N_2^2 - \xi N_2 N_1 + q_2(t) \tag{6}$$

Each random function represents the set of events affecting the population of a specie. It includes randomness in growth rate, fluctuations in habitat conditions (environmental stochasticity), random changes in reproductive rates, etc. Each of these independent random processes may have its own statistical distribution and correlation time. However, by invoking the Central Limit Theorem, we can assume their sum total tends towards a single Gaussian distribution. This assumption is valid as long as there is not a dominating and high impact random event amongst the set. We further assume that the resulting distribution is zero mean and $\delta$- correlated in time. Thus,

$$<q_i(t)> = 0 \quad , \quad <q_i^*(t) q_j(t')> = 2\delta_{ij}\delta(t - t') \quad i, j = 1, 2 \tag{7}$$

where $\delta_{ij}$ is the Kronecker delta and $\delta(t-t')$ is the Dirac delta function. $\delta_{ij}$ represents our assumption of statistical independence of $q_1(t)$ and $q_2(t)$. This assumption is not strictly valid as events such as environmental stochasticity that affects a common geographical area containing the two species may influence both populations in a correlated fashion. Nevertheless, the assumption is a good approximation as long as the correlated events are not the dominant events in the ensemble of random processes or their amplitudes are small compared the two species' deterministic growth rates.

The zero-mean assumption ($<q_i(t)>=0$) besides being a mathematical convenience indicates that in the totality of random processes affecting each specie's population some have positive and some have negative effects on the population. If there is a bias in the effects it can be included in the (deterministic) growth rate terms of Eqs. (4) and (5).

The Dirac delta function in (7) indicates our assumption that our Gaussian distribution has a white noise spectrum. This assumption is valid only if the correlation times of the underlying random processes are much smaller than the system's macroscopic time. The latter is defined as the time span of the macroscopic evolution of the system when subjected to a random event. It can be considered as the system's relaxation time towards its steady state. The white noise is a good approximation since, in general, the environmental state varies much faster the system's macroscopic sate [14].

Eqs. (5) and (6) are known as Langevin equations of the random processes $N_1$ and $N_2$. As is well known the set can be transformed to a Fokker-Planck equation [15] whose general form is:

$$\frac{\partial}{\partial t}P = -\sum_{i=1}^{2}\frac{\partial}{\partial N_i}(A_i P) + \frac{1}{2}\sum_{i,j=1}^{2}D_{ij}\frac{\partial^2}{\partial N_i \partial N_j}P \qquad (8)$$

where $P \equiv P(N_1, N_2, t)$ is the joint probability density of the two species having populations $N_1$ and $N_2$ at time $t$, $A_1$ and $A_2$ are components of the "drift vector" and $D_{ij}$ is the diffusion matrix. For the Langevin equations (4) and (5) with the conditions (7) the drift vector and diffusion matrix are given as:

$$A_1 = (a_1 - N_1 - \xi N_2)N_1 \quad , \quad A_2 = (a_2 - N_2 - \xi N_1)N_2 \quad , \quad D_{ij} = 2\delta_{ij} \qquad (9)$$

## IV. Steady-State Solution of Joint Probability Density

After a sufficiently long time, the two-species system reaches a steady state in which the joint probability density becomes time independent. The steady state solution to the Fokker-Planck equation (8) with components of drift vector and diffusion coefficient given by (9) can be shown to be [12]:

$$P(N_1, N_2) = Q^{-1}\exp\left(\frac{1}{2}a_1 N_1 + \frac{1}{2}a_2 N_2 - \frac{1}{4}N_1^2 - \frac{1}{4}N_2^2 - \frac{1}{2}\xi N_1 N_2\right) \qquad (10)$$

where $Q$ is the normalization constant determined from the condition:

$$\int_0^\infty\int_0^\infty P(N_1,N_2)dN_1dN_2 = 1 \tag{11}$$

and expressed as:

$$Q(a_1,a_2,\xi) = \int_0^\infty\int_0^\infty \exp\left(\frac{1}{2}a_1N_1 + \frac{1}{2}a_2N_2 - \frac{1}{4}N_1^2 - \frac{1}{4}N_2^2 - \frac{1}{2}\xi N_1 N_2\right)dN_1 dN_2 \tag{12}$$

The double integral in (12) can be simplified further by completing the square with respect to $N_2$ in the exponent and then integrating over $N_2$. We obtain

$$Q(a_1,a_2,\xi) = 2\sqrt{\frac{\pi}{1-\xi^2}}\exp(b_1^2 + \frac{1}{4}a_2^2)\int_{-b_1}^\infty dy \exp(-y^2) erfc\left(\frac{\xi y - b_2}{\sqrt{1-\xi^2}}\right) \tag{13}$$

where $erfc(x)$ is the complementary error function defined by

$$erfc(x) \equiv 1 - erf(x) = 1 - \frac{2}{\sqrt{\pi}}\int_0^x e^{-t^2}dt$$

with the $b_1$, $b_2$ parameters defined as

$$b_1 = \frac{a_1 - \xi a_2}{2\sqrt{1-\xi^2}}, \qquad b_2 = \frac{a_2 - \xi a_1}{2\sqrt{1-\xi^2}} \tag{14}$$

In the above transformation, we have assumed $\xi \neq 1$. The case of $\xi=1$ will be treated separately in Section VI.2

The behavior of the joint probability density (10) depends strongly on the growth rates of the two species. Far below the threshold where both $a_1$ and $a_2$ are negative and numerically large, the terms $\frac{1}{2}a_1N_1$ and $\frac{1}{2}a_2N_2$ in the exponent dominate, and to a good approximation, we may write

$$P(N_1,N_2) \approx \frac{1}{4}|a_1 a_2|\exp[-\frac{1}{2}(|a_1|N_1 + |a_1|N_2)]$$

which is a distribution characteristic of two independent populations each with a Poisson distribution and means of $2/|a_1|$ and $2/|a_2|$ and standard deviations equal to their respective mean . In this case, competition has little effect on population dynamics.

In the opposite limit of far above thresholds ($a_1, a_2 \gg 1$), it can be shown [12] that the joint probability density (9) becomes genuinely Gaussian to a very good approximation and the means, the variances, and the covariance of $N_1$ and $N_2$ can be written as [16]:

$$<N_1> \approx \frac{a_1 - \xi a_2}{1 - \xi^2} \quad , \quad <N_2> \approx \frac{a_2 - \xi a_1}{1 - \xi^2} \tag{15a}$$

$$\frac{<(\Delta N_1)^2>}{<N_1>^2} \approx \frac{<(\Delta N_2)^2>}{<N_2>^2} \approx \frac{2}{1-\xi^2} \tag{15b}$$

$$\frac{<\Delta N_1 \Delta N_2>}{\sqrt{<(\Delta N_1)^2><(\Delta N_2)^2>}} \approx -\xi \tag{15c}$$

$$\frac{<\Delta N_1 \Delta N_2>}{<N_1><N_2>} \approx \frac{-2\xi(1-\xi^2)}{(a_1 - \xi a_2)(a_2 - \xi a_1)} \tag{15d}$$

where we have written $\Delta N_1 = N_1 - <N_1>$ and $\Delta N_2 = N_2 - <N_2>$.

Eqs. (15) imply that in the limit of $a_1, a_2 \gg 1$ the population means approach those of the steady state in the absence of random events (see Eqs. 3) with relative fluctuations completely determined by the competition coefficient. Furthermore, the population fluctuations of the two species become anti-correlated with a correlation coefficient that is numerically equal to the competition coefficient $\xi$. As expected, in this limit, an increase in one population is accompanied by a decrease in the other with a ratio determined by the competition coefficient.

## V. Moments of Distribution

The normalization parameter $Q(a_1, a_2, \xi)$ also serves as the generating function for the moments of species populations. Thus, the population mean associated with specie $s$ ($s = 1, 2$) is given by

$$<N_s> = \int_0^\infty \int_0^\infty N_s P(N_1, N_2) dN_1 dN_2 = \frac{2}{Q} \frac{\partial Q}{\partial a_s} = 2 \frac{\partial}{\partial a_s}(\ln Q) \tag{16}$$

and, more generally, the moments of order $n + m$ can be obtained from

$$<N_s^n N_{s'}^m> = \int_0^\infty \int_0^\infty N_s^n N_{s'}^m P(N_1, N_2) dN_1 dN_2 = \frac{2^{n+m}}{Q} \frac{\partial^{n+m} Q}{\partial a_s \partial a_{s'}} \tag{17}$$

We are more interested in the moments of the population fluctuations about the mean, i.e. $\Delta N_s \equiv N_s - <N_s>$. The second order moment takes the simple form of

$$M_{ss'} \equiv <\Delta N_s \Delta N_{s'}> = <N_s N_{s'}> - <N_s><N_{s'}> = \frac{4}{Q}\frac{\partial^2 Q}{\partial a_s \partial a_{s'}} - \frac{4}{Q^2}\frac{\partial Q}{\partial a_s}\frac{\partial Q}{\partial a_{s'}}$$

$$= 2\frac{\partial <N_{s'}>}{\partial a_s} = 2\frac{\partial <N_s>}{\partial a_{s'}} \qquad s, s' = 1, 2 \qquad (18)$$

Using (13) in (16) and (18) we find the following expressions for the population means and fluctuations of the two species [12]:

$$<N_1(a_1, a_2, \xi)> = \frac{a_1 - \xi a_2}{1-\xi^2} + \frac{2\sqrt{\pi}}{(1-\xi^2)Q}\left\{\exp(\frac{1}{4}a_2^2)\left[1+erf(\frac{1}{2}a_2)\right] - \xi\exp(\frac{1}{4}a_1^2)\left[1+erf(\frac{1}{2}a_1)\right]\right\} \quad (19)$$

$$<N_2(a_1, a_2, \xi)> = \frac{a_2 - \xi a_1}{1-\xi^2} + \frac{2\sqrt{\pi}}{(1-\xi^2)Q}\left\{\exp(\frac{1}{4}a_1^2)\left[1+erf(\frac{1}{2}a_1)\right] - \xi\exp(\frac{1}{4}a_2^2)\left[1+erf(\frac{1}{2}a_2)\right]\right\} \quad (20)$$

$$<(\Delta N_1)^2> = \frac{2}{1-\xi^2} - \left(<N_1> - \frac{a_1 - \xi a_2}{1-\xi^2}\right)^2 - \frac{2\sqrt{\pi}}{(1-\xi^2)^2 Q}$$

$$\times\left\{\exp(\frac{1}{4}a_2^2)(a_1 - \xi a_2)[1+erf(\frac{1}{2}a_2)] + \xi^2\exp(\frac{1}{4}a_1^2)(a_2 - \xi a_1)[1+erf(\frac{1}{2}a_1)] + \frac{2\xi(1-\xi^2)}{\sqrt{\pi}}\right\} \quad (21)$$

$$<(\Delta N_2)^2> = \frac{2}{1-\xi^2} - \left(<N_2> - \frac{a_2 - \xi a_1}{1-\xi^2}\right)^2 - \frac{2\sqrt{\pi}}{(1-\xi^2)^2 Q}$$

$$\times\left\{\exp(\frac{1}{4}a_1^2)(a_2 - \xi a_1)[1+erf(\frac{1}{2}a_1)] + \xi^2\exp(\frac{1}{4}a_2^2)(a_1 - \xi a_2)[1+erf(\frac{1}{2}a_2)] + \frac{2\xi(1-\xi^2)}{\sqrt{\pi}}\right\} \quad (22)$$

$$<\Delta N_1 \Delta N_2> = -\frac{2\xi}{1-\xi^2} - \left(<N_1> - \frac{a_1 - \xi a_2}{1-\xi^2}\right)\left(<N_2> - \frac{a_2 - \xi a_1}{1-\xi^2}\right) + \frac{2\sqrt{\pi}\xi}{(1-\xi^2)^2 Q}$$

$$\times\left\{\exp(\frac{1}{4}a_1^2)(a_2 - \xi a_1)[1+erf(\frac{1}{2}a_1)] + \exp(\frac{1}{4}a_2^2)(a_1 - \xi a_2)[1+erf(\frac{1}{2}a_2)] + \frac{2(1-\xi^2)}{\sqrt{\pi}}\right\} \quad (23)$$

The below diagrams show plots of Eqs.(19)-(23) for several combinations of $a_1, a_2$, and $\xi$. In each graph the abscissa is the growth rate $a_1$ assumed to be that of the stronger specie and the third axis is a new parameter $\delta$ that is related to $\xi$ through

$$\xi = \frac{1}{1+\delta^2} \qquad (24)$$

Also, in each diagram, the difference of growth rates $\Delta a = a_1 - a_2$ is held constant. The diagrams illustrate the behavior of the two populations and their fluctuations and correlations for $\Delta a$ having the values of 0 and 5 that represent two species of equal and very unequal growth rates, respectively. The parameter $\delta$ varies between 0 (corresponding to maximum competition, $\xi = 1$) and 2 (corresponding to little competition, $\xi = 0.2$)

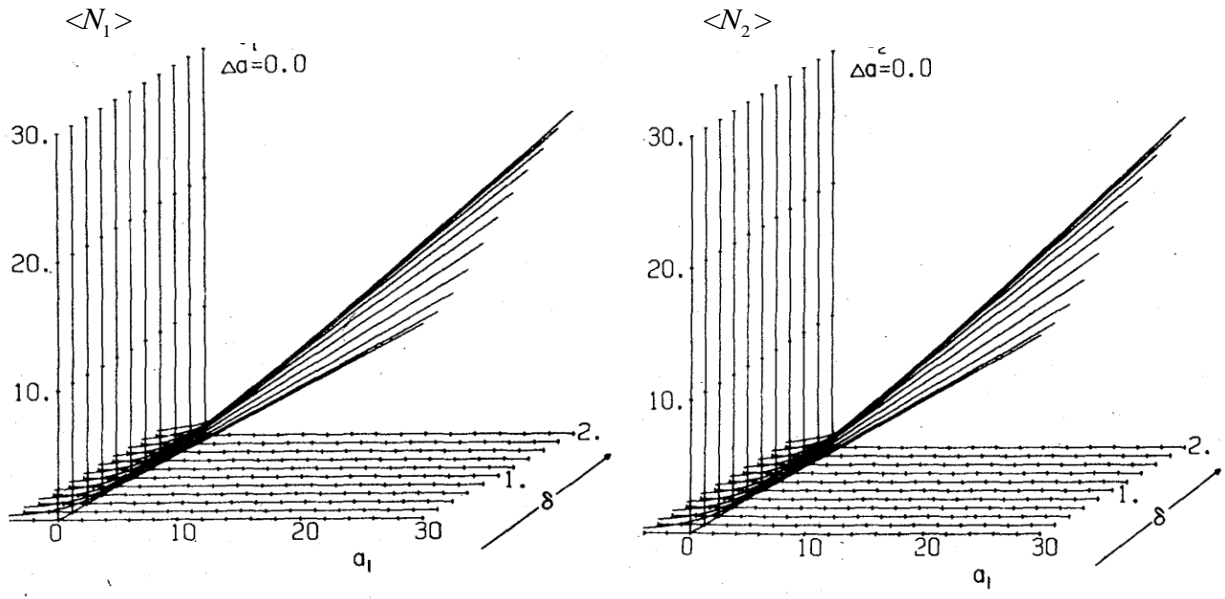

Fig. 1. Population means of species 1 and 2 as functions of their (equal) growth rates for various degrees of competition.

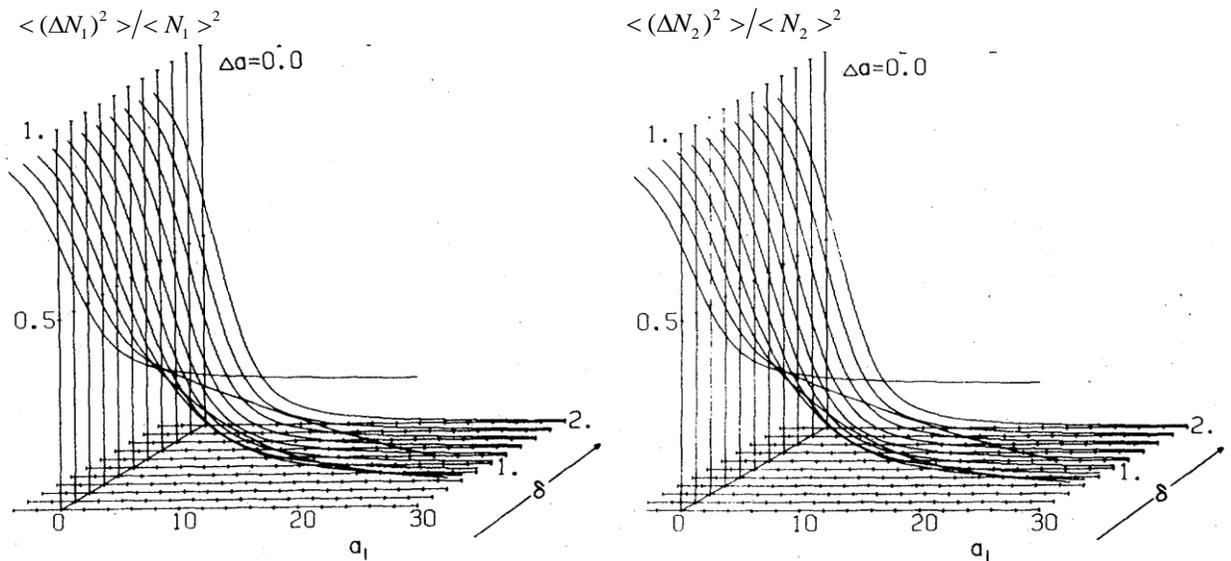

Fig. 2. Normalized population fluctuations of species (1) and (2) as functions of their (equal) growth rates for various degrees of competition.

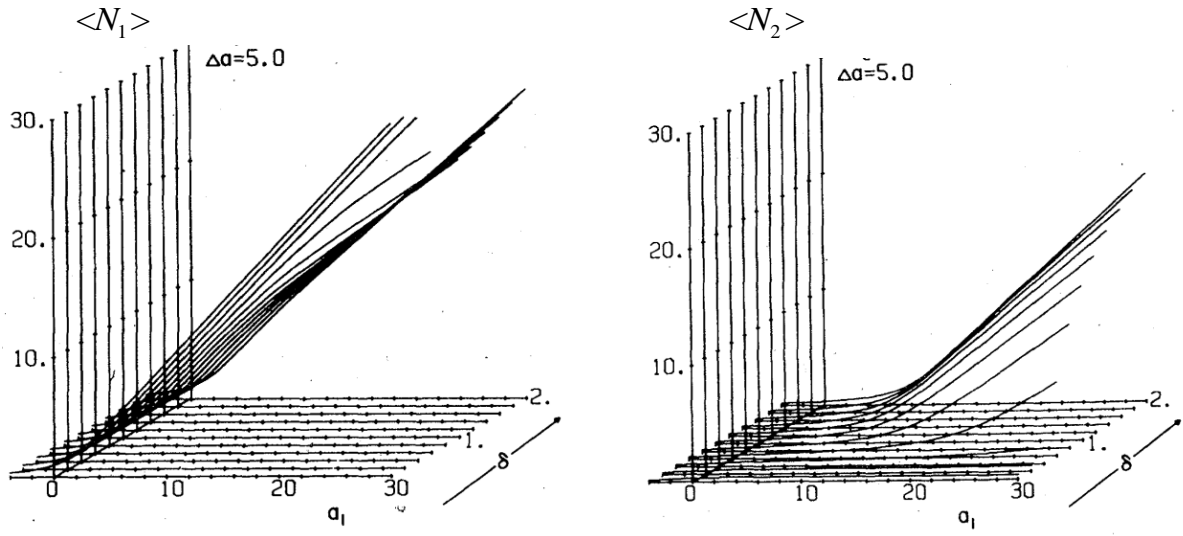

Fig. 3. Population means of species 1 and 2 as functions of their (unequal) growth rates for various degrees of competition.

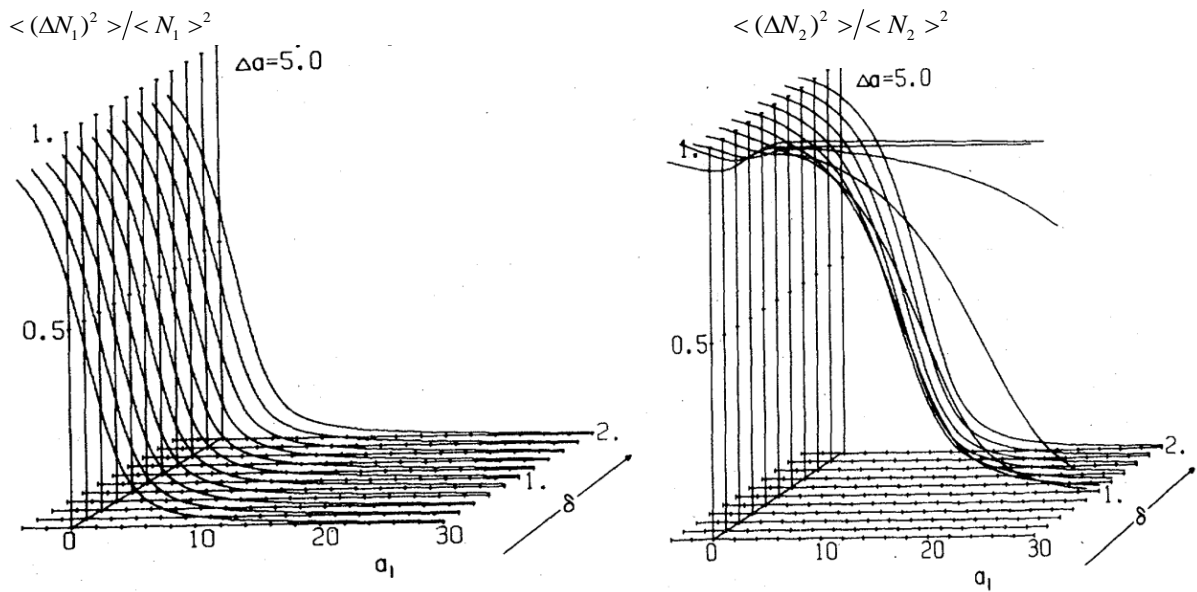

Fig. 4. Normalized population fluctuations of species (1) and (2) as functions of their (unequal) growth rates for various degrees of competition.

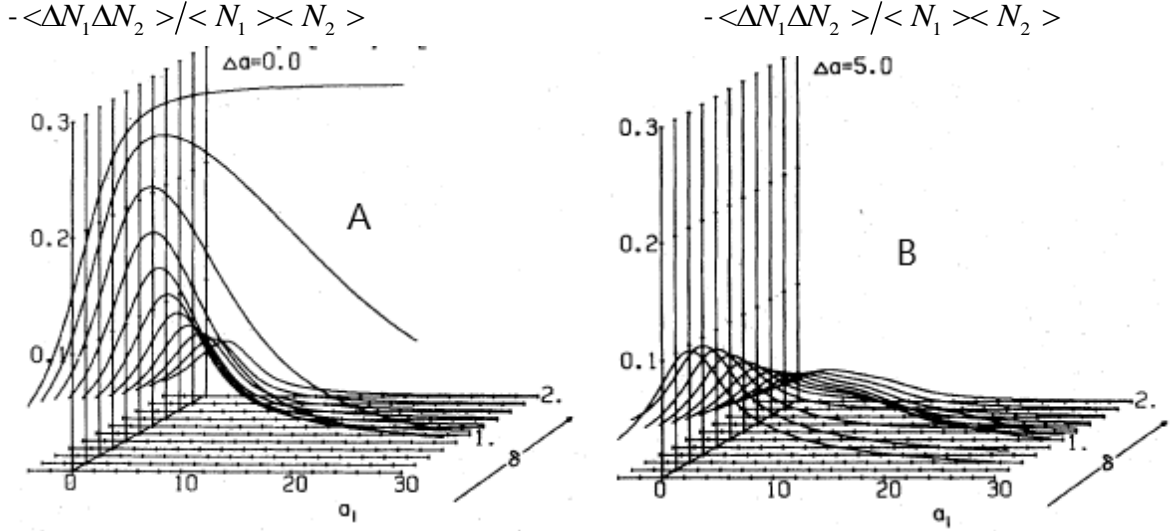

Fig. 5. Negative of the normalized cross correlations of the two species populations for various degrees of competition. A: for equal ($\Delta a = 0$) and B: for unequal ($\Delta a = 5$) growth rates.

Some of the noticeable features apparent in these figures are:

a) For equal growth rates, the two species population means reach the same level (Fig.1). However, with increasing competition the normalized fluctuations in each population do not die out. In fact, for $\xi = 1$ the relative fluctuations approach a constant value of 1/3 regardless of how large the growth rates are (Figs.2). Thus, maximum competition generates fluctuations in both populations with a standard deviation of $1/\sqrt{3}$ of the population mean. At the same time the normalized cross correlation reaches a value of -1/3 (Fig.5A)

b) For unequal growth rates, the population mean of the weaker specie ($<N_2>$) fails to grow with its increasing growth rate ($a_2$) when the difference of growth rates become significant (Fig. 3). While the normalized fluctuations of the stronger specie quickly die out as its growth rate increases those of the weaker specie do not. This behavior becomes more striking as the competition increases. In fact, for $\xi = 1$ the weaker specie's normalized fluctuations reach 1 regardless of how large its growth rate becomes (Fig.4). It should be noted that for the weaker specie even though $<N_2>$ does not grow with $a_2$ it obtains a finite, non-zero, value. For $\xi = 1$ where normalized fluctuations reach 1, the standard deviation becomes equal to the population mean. In other words, under maximum competition, the weaker specie never becomes extinct but attains a population mean that fluctuates with a standard deviation equal to the mean.

The above features are all reflections of the competition between the two species for the same resource in an environment influenced by random events and they become most striking for $\xi = 1$ where the competition is greatest. This situation needs special attention for which Eqs. (19)-(23) can be evaluated exactly. We will come back to this point in Section VI.2.

# VI. Special Cases

There are several limits of Eq. (19)-(23) that require special attention. The cases considered reveal the system behavior at the growth rate threshold as well as under the extremes of competition.

## VI.1. Two Species at Threshold

The population threshold is characterized by the vanishing of growth rate. In a two-species system it is of course possible that the population of one specie be at threshold while the other one is not. However, the case of both species population at threshold is an interesting one and it provides a drastic simplification of the equations for population means and correlation functions.

For $a_1 = a_2 = 0$, the generating function $Q(0,0,\xi)$ becomes a function of $\xi$ only and the integral in Eq. (12) can be evaluated exactly to obtain

$$Q(0,0,\xi) = \frac{2\sqrt{\pi}}{\sqrt{1-\xi^2}} \int_0^\infty e^{-t^2} erfc(\frac{\xi t}{\sqrt{1-\xi^2}}) dt = \frac{2}{\sqrt{1-\xi^2}} \arccos \xi . \qquad (44)$$

In this case, Eqs. (17) and (18) for average populations reduce to:

$$<N_s(0,0,\xi)> = \sqrt{\pi \frac{1-\xi}{1+\xi}} \frac{1}{\arccos \xi} \qquad s=1,2 \qquad (45)$$

and the expressions for relative fluctuations, Eqs. (19) and (20) become

$$\frac{<[\Delta N_s(0,0,\xi)]^2>}{<N_s(0,0,\xi)>^2} = \frac{2(\arccos \xi)^2}{\pi(1-\xi)^2} - \frac{2\xi(1+\xi)\arccos \xi}{\pi(1-\xi)\sqrt{1-\xi^2}} - 1 \qquad s=1,2 \qquad (46)$$

Finally, from Eq. (21) we obtain the following expression for the relative cross correlations of the population fluctuations at threshold,

$$\frac{<\Delta N_1(0,0,\xi)\Delta N_2(0,0,\xi)>}{<N_1(0,0,\xi)><N_2(0,0,\xi)>} = \frac{2\arccos \xi}{\pi(1-\xi^2)^{3/2}} - \frac{2\xi^2(\arccos \xi)^2}{\pi(1-\xi^2)^2} - 1 \qquad (47)$$

We observe that as the competition approaches its maximum ($\xi \to 1$), Eqs. (45)-(47) reduce to

$$<N_s(0,0,1)> \to \frac{\sqrt{\pi}}{2} \qquad (48)$$

$$\frac{<[\Delta N_s(0,0,1)]^2>}{<N_s(0,0,1)>^2} \to \frac{4}{\pi} - 1 \qquad (49)$$

$$\frac{<\Delta N_1(0,0,1)\Delta N_2(0,0,1)>}{<N_1(0,0,1)><N_2(0,0,1)>} \to \frac{2}{\pi}-1 \qquad (50)$$

where we have used the approximation $\arccos \xi \to \sqrt{1-\xi^2}$ as $\xi \to 1$. As expected, Eqs.(48)-(50) are consistent with Eqs. (31)-(33) for $a=0$.

## VI.2. Two Species at Maximum Competition

The case of maximum competition is especially interesting as the statistical fluctuations of species populations play a dominant role. As noted earlier, for $\xi=1$ and in the absence of random Langevin events $q_1(t)$, $q_2(t)$, the coupled equations (1) and (2) have no steady-state solutions in which both populations are nonzero unless the growth rates are equal. When $\xi=1$ the integral in (13) can be calculated explicitly and leads to exact solutions for Eqs. (18)-(22). We find [12]:

$$Q(a_1,a_2,1) = \frac{2\sqrt{\pi}}{a_1-a_2}\left\{\exp(\frac{1}{4}a_1^2)\left[1+erf(\frac{1}{2}a_1)\right] - \exp(\frac{1}{4}a_2^2)\left[1+erf(\frac{1}{2}a_2)\right]\right\} \qquad (24)$$

$$<N_1(a_1,a_2,1)> = \frac{2}{a_2-a_1} + \frac{a_1(\exp(\frac{1}{4}a_1^2)[1+erf(\frac{1}{2}a_1)] + \frac{2}{\sqrt{\pi}}}{\exp(\frac{1}{4}a_1^2)[1+erf(\frac{1}{2}a_1)] - \exp(\frac{1}{4}a_2^2)[1+erf(\frac{1}{2}a_2)]} \qquad (25)$$

$$<N_2(a_1,a_2,1)> = \frac{2}{a_1-a_2} + \frac{a_2(\exp(\frac{1}{4}a_2^2)[1+erf(\frac{1}{2}a_2)] + \frac{2}{\sqrt{\pi}}}{\exp(\frac{1}{4}a_2^2)[1+erf(\frac{1}{2}a_2)] - \exp(\frac{1}{4}a_1^2)[1+erf(\frac{1}{2}a_1)]} \qquad (26)$$

$$<(\Delta N_1)^2> = \frac{4}{(a_1-a_2)^2} - \left(<N_1(a_1,a_2,\xi)> - \frac{2}{a_2-a_1}\right)^2$$
$$+ \frac{(2+a_1^2)(\exp(\frac{1}{4}a_1^2)[1+erf(\frac{1}{2}a_1)] + \frac{2a_1}{\sqrt{\pi}}}{\exp(\frac{1}{4}a_1^2)[1+erf(\frac{1}{2}a_1)] - \exp(\frac{1}{4}a_2^2)[1+erf(\frac{1}{2}a_2)]} \qquad (27)$$

$$<(\Delta N_2)^2> = \frac{4}{(a_2-a_1)^2} - \left(<N_2(a_1,a_2,\xi)> - \frac{2}{a_1-a_2}\right)^2$$
$$+ \frac{(2+a_2^2)(\exp(\frac{1}{4}a_2^2)[1+erf(\frac{1}{2}a_2)] + \frac{2a_2}{\sqrt{\pi}}}{\exp(\frac{1}{4}a_2^2)[1+erf(\frac{1}{2}a_2)] - \exp(\frac{1}{4}a_1^2)[1+erf(\frac{1}{2}a_1)]} \qquad (28)$$

$$<\Delta N_1 \Delta N_2> = \frac{2(<N_1> - <N_2>)}{(a_1 - a_2)} - <N_1><N_2> \quad (29)$$

We observe that the case of equal growth rates requires special attention. Using $a_1 = a_2 \equiv a$ in Eqs. (24)-(29) we obtain

$$Q(a,a,1) = \sqrt{\pi}\{ae\,xp(\frac{1}{4}a^2)[1+erf(\frac{1}{2}a)] + \frac{2}{\sqrt{\pi}}\} \quad (30)$$

$$<N_1(a,a,1)> = <N_2(a,a,1)> = \frac{1}{2}a + \frac{1+erf(\frac{1}{2}a)}{a[1+erf(\frac{1}{2}a)] + \frac{2}{\sqrt{\pi}}\exp(-\frac{1}{4}a^2)} \quad (31)$$

$$<(\Delta N_1)^2> = <(\Delta N_2)^2> = 1 + (\frac{2}{3}a - <N_1>)<N_2> \quad (32)$$

$$<\Delta N_1 \Delta N_2> = \frac{1}{2} + (\frac{1}{3}a - <N_1>)<N_2> \quad (33)$$

The above equations have some interesting consequences for the behavior of the two-species system when their (equal) growth rates are well above threshold. For, when $a \gg 1$ Eqs. (31)-(33) reduce to:

$$<N_1(a,a,1)> = <N_2(a,a,1)> \Rightarrow \frac{1}{2}a \quad (34)$$

$$\frac{<(\Delta N_1)^2>}{<N_1>^2} = \frac{<(\Delta N_2)^2>}{<N_2>^2} \Rightarrow \frac{1}{3} \quad (35)$$

$$\frac{<\Delta N_1 \Delta N_2>}{<N_1><N_2>} \Rightarrow -\frac{1}{3} \quad (36)$$

$$\frac{<\Delta N_1 \Delta N_2>}{\sqrt{<(\Delta N_1)^2><(\Delta N_2)^2>}} \Rightarrow -1 \quad (37)$$

Hence, for equal growth rates at maximum competition, normalized population fluctuations do not die out well above the threshold and approach the constant value of 1/3. Also, as shown in (37), populations of the two species become completely anti-correlated.

In general, one may show that when the growth rates of the two species are unequal, population of the weaker species becomes more and more random even as its growth rate increases. Let us assume that $\Delta a = a_1 - a_2 > 0$ and that $\Delta a$ is held constant as $a_1 \to \infty$. In other words, while the growth rates of both species are increasing, one (the weaker specie) is always less than the other (the stronger specie) by a constant amount. In this situation, we find from Eqs. (25) and (26),

$$<N_1(a_1,a_1-\Delta a,1)> \;\to a_1-2/\Delta a \tag{38}$$

$$<N_2(a_1,a_1-\Delta a,1)> \;\to 2/\Delta a \tag{39}$$

so that the population of the weaker of the two species does not grow with its increasing growth rate, but tend towards a constant value. This is a reflection of the competition for the same resource between the two species. Also, in the same scenario, we obtain from Eqs. (27) and (28),

$$<[\Delta N_1(a_1,a_1-\Delta a,1)]^2> \;\to 2+4/(\Delta a)^2 \tag{40}$$

$$<[\Delta N_2(a_1,a_1-\Delta a,1)]^2> \;\to 4/(\Delta a)^2 \tag{41}$$

so that

$$\frac{<(\Delta N_1)^2>}{<N_1>^2} \to \frac{2+4/(\Delta a)^2}{a_1^2} \quad \text{and} \quad \frac{<(\Delta N_2)^2>}{<N_2>^2} \to 1 \tag{42}$$

We see that whereas the relative fluctuations of the stronger specie gradually die out well above the threshold, those of the weaker specie do not. In fact, the relative population fluctuations of the weaker specie tend towards the complete chaos. Thus, in maximum competition, the weaker specie does not necessarily become extinct, rather, its population statistics becomes fully chaotic. From (39) and (41) we find that the weaker specie's population will attain a mean value of $2/\Delta a$ with a standard deviation of $2/\Delta a$.

The cross correlation between the population fluctuation of the two species can be found from (28), (37), and (38). We obtain

$$\frac{<\Delta N_1 \Delta N_2>}{<N_1><N_2>} \to -\frac{2}{a_1 \Delta a} \tag{42}$$

and

$$\frac{<\Delta N_1 \Delta N_2>}{\sqrt{<(\Delta N_1)^2><(\Delta N_2)^2>}} \to -\frac{2}{\Delta a\sqrt{2+4/(\Delta a)^2}} \tag{43}$$

which, when considered along with (41), show that even though the relative fluctuations of the stronger specie tend to die out above threshold, the cross correlation coefficient between the population fluctuations remains constant. Most of these features are illustrated by the graphs in Figs. 1-5.

## VI.3. Two Non-Interacting Species

Finally, we consider the case of two non-interacting species for which $\xi = 0$. This case describes the population dynamics of a single specie under the influence of random events.

For $\xi = 0$ the generating function shown in Eq. (12) can be calculated exactly and we find

$$Q(a_1, a_2, 0) = \pi \exp(\frac{1}{4}a_1^2)[1+\text{erf}(\frac{a_1}{2})]\exp(\frac{1}{4}a_2^2)[1+\text{erf}(\frac{a_2}{2})] \qquad (51)$$

whose substitution in Eqs. (18) and (20) yields

$$<N_1(a_1,a_2,0)> = a_1 + \frac{2}{\sqrt{\pi}} \frac{\exp(\frac{-1}{4}a_1^2)}{1+\text{erf}(\frac{a_1}{2})} \qquad (52)$$

and

$$<[\Delta N_1(a_1,a_2,0)]^2> = 2 - [<N_1(a_1,a_2,0)> - a_1]^2 - \frac{2a_1}{\sqrt{\pi}} \frac{\exp(\frac{-1}{4}a_1^2)}{1+\text{erf}(\frac{a_1}{2})} \qquad (53)$$

with identical expressions for the second specie with the substitution of $a_2$ for $a_1$. As expected, all references to the other specie in the population mean and population fluctuations of one specie have disappeared. The situation is reminiscent of a single-mode laser under the influence of random processes and for which identical expressions have been derived [15].

The first term on the right hand side of Eq. (52) shows the steady state solution to Lokta-Volterra Eq. (3) for a single specie. The second term shows the effect of random events. It is interesting to note that with the first term only, the population mean is driven to extinction at growth rate threshold ($a_1 = 0$). However, the inclusion of random events provides a non-zero value at threshold and leads to a smooth transition from far above threshold ($a_1 \gg 0$) to far below threshold ($a_1 \ll 0$), as shown in Fig. (6). This indicates that random events may save the specie from extinction at or below its growth rate threshold.

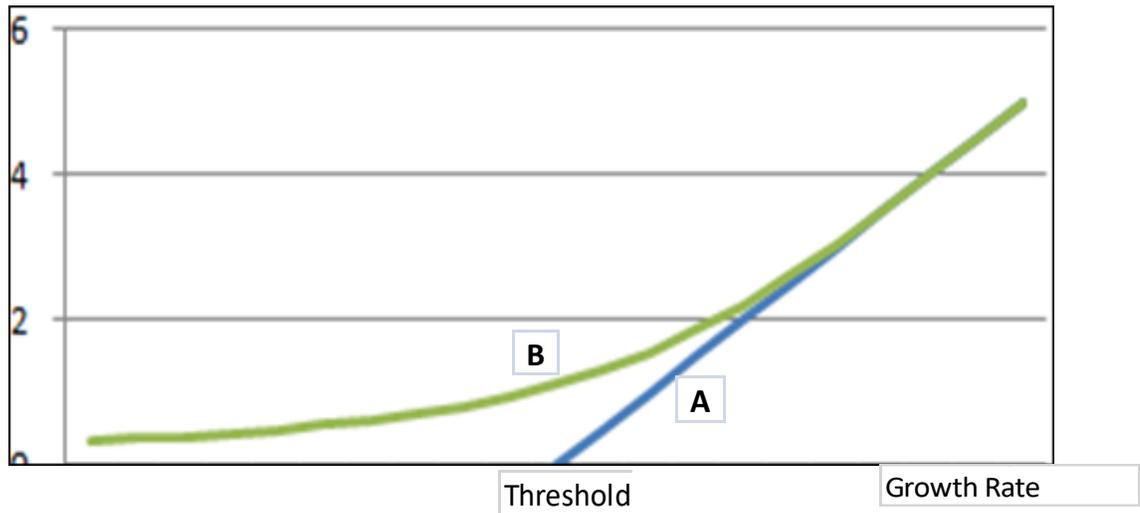

Fig. 6. Population mean of a non-interacting single specie vs. its growth rate. The straight line A shows the population without the influence of random events. The curved line B shows the effect of random events.

It is interesting to compare the population mean of a single specie, Eq.(52), with that of two species with equal growth rates under maximum competition, Eq. (31). A plot of the latter would be identical to Fig.6 but with different values on the y-axis. A combined plot of the two cases is shown below:

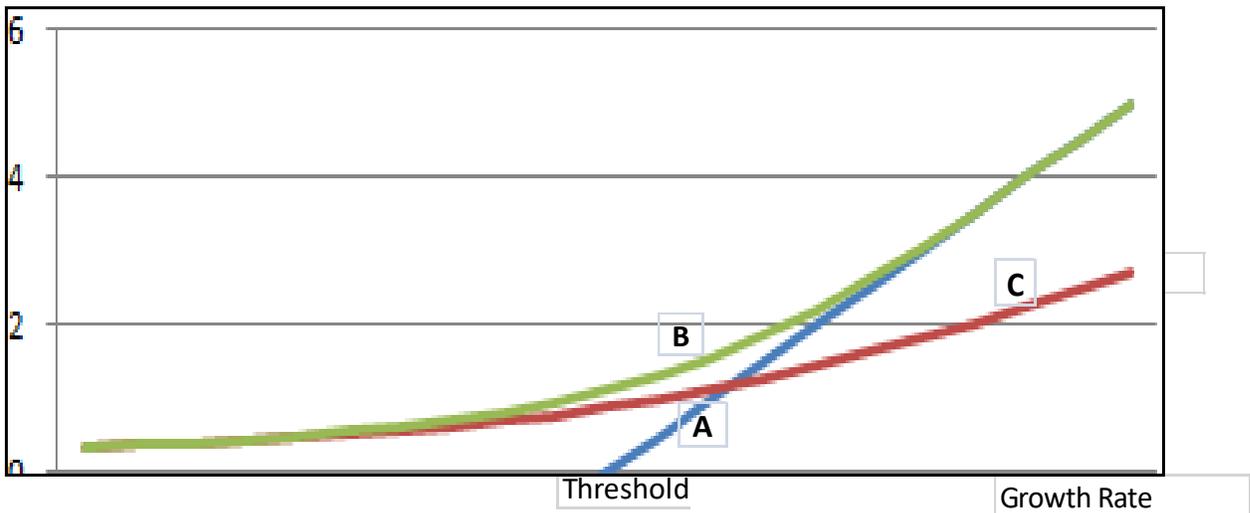

Fig. 7. Comparison of the population means of a single specie and an equal strength two-species system under maximum competition (curve C). Curves A and B are the same as in Fig. 6.

We note from Eq. (4) that for equal growth rates and in the absence of random events the population of two species at maximum competition follows a straight line ($a/2$) and becomes zero at threshold. However, inclusion of random events results in a population mean for both species that follows curve C in Fig. 7 and provides a smooth transition from far above to far below threshold. Eq. (31) indicates that at threshold the scaled population mean of both species is equal to $\sqrt{\pi}/2$. In situations where the competition coefficient is between 0 and 1 the population mean as a function of growth rate falls at a point in the area between the curves B and C.

We should emphasize that the transition in population mean from far above to far below the growth rate is accompanied by increasing population fluctuations. In the above example, while the scaled population mean is equal to $\sqrt{\pi}/2 = 0.89$, its standard deviation, derived from Eq.(32), is equal to 0.46. Similarly, while the scaled population mean of a single specie at threshold, according to Eq.(52), is equal to $2/\sqrt{\pi} \approx 1.13$ its standard deviation, according to Eq.(53), is equal to 0.85. The scale factor in these examples is determined by the parameters used in setting up the Langevin equations (4) and (5).

# VII. Summary and Conclusions

In this paper we have used the behavior of a two-mode ring laser as an analog of two competing species under the influence of random events. We have shown that, with the exception of the nature of the system parameters, the two can be described by an identical set of equations. We discussed the differences between the two systems' parameters and laid out the assumptions and conditions under which the identical formulation is valid. We have set up the Fokker–Planck equation of the two-species system and obtained the steady-state solution for the probability density function and used it to obtain expressions for the population mean and fluctuations of each specie. We have shown that competition increases the population fluctuations of each specie. In fact, under maximum competition and for equal growth rates the relative fluctuations of both species populations never die out and reach a constant value of 1/3 regardless of how large their growth rates are. For unequal growth rates the population mean of the weaker specie fails to grow with its growth rate and, while the normalized fluctuations of the stronger specie diminish as its growth rate increases, those of the weaker specie do not. And, under maximum competition, the weaker specie's normalized fluctuations reaches 1 regardless of how large its growth rate becomes. We have shown that, when random events are taken into account, the population mean transition from high above (growth rate) threshold to far below threshold follows a smooth curve and always stays positive. In other words, random events save the weaker specie from complete extinction. At the same time, the population fluctuations increase as the system goes through the threshold and below. The underlying assumption is that in all the random processes affecting the species none is dominant over all others.

We have considered the competition parameter, $\xi$, to be positive to indicate competition. The case of $\xi$ <0 relates to cooperative entities that we have not considered here. Clearly, the mathematical formulation of cooperating entities is identical to what is presented here but the results will have different interpretations. In this regard, it is worth emphasizing that there exist numerous systems which can be described through competitive or cooperative interactions. To mention just a few: biological assemblies of individuals, coupled chemical reactions, political parties, businesses, and countries [5]. Thus, our results are not restricted only to population dynamics of species, but the main ideas can be applied to a wide variety of situations embracing different scientific areas.

Finally, this paper does not address the time evolution of the dynamics of the two-entity system. That requires the time dependent solution to the Fokker-Planck equation (8). While the time independent solution of Eq. (8) allowed us to derive mathematical expressions for the steady or "equilibrium" state of the system, the time dependent solution will provide, among other things, formulas for the correlation times of the system to which the random events' periods can be compared. This topic will be the subject of a future publication.